\documentclass[a4paper]{article}

\usepackage[utf8]{inputenc} 
\usepackage{hyperref}       
\usepackage{url}            
\usepackage{amsfonts}       
\usepackage{amsmath}
\usepackage{graphicx}
\usepackage{caption}
\usepackage{float}
\usepackage[left=3cm,right=3cm,top=2cm,bottom=2cm]{geometry}
\usepackage[english]{babel}
\usepackage{amsmath}
\usepackage{graphicx}
\usepackage[colorinlistoftodos]{todonotes}
\usepackage{float}
\usepackage{dsfont}
\usepackage{bm} 
\usepackage{url}
\usepackage{booktabs} 
\usepackage{multicol}
\usepackage{multirow}
\usepackage{tabularx}
\DeclareCaptionFormat{sanslabel}{#3}%
\usepackage{epstopdf} 
\usepackage{dblfloatfix}

\makeatletter
\def\@seccntformat#1{\@ifundefined{#1@cntformat}%
   {\csname the#1\endcsname\quad}
   {\csname #1@cntformat\endcsname}}
\newcommand\section@cntformat{}     
\makeatother

\begin{document}

\Large
\begin{center}
\vspace{1cm}






\textbf{Curved-Full-Field OCT for high-resolution imaging of living human retina over a large field-of-view}

\vspace{0.5cm}

\normalsize
Pedro Mecê$^{1,*}$, Kassandra Groux$^1$, Jules Scholler$^{1}$,  Olivier Thouvenin$^{1}$, Mathias Fink$^1$, Kate Grieve$^{2,3}$ and Claude Boccara$^1$

\end{center}
\scriptsize
$^1$Institut Langevin, CNRS, ESPCI Paris, PSL Research University, 10 rue Vauquelin, Paris, France\\
$^2$Institut de la Vision, Sorbonne Université, INSERM, CNRS, F-75012, Paris, France\\
$^3$Quinze-Vingts National Eye Hospital, 28 Rue de Charenton, Paris, 75012, France\\
$^*$pedro.mece@espci.fr

\normalsize


\begin{abstract} 

Allying high-resolution with a large field-of-view (FOV) is of great importance in the fields of biology and medicine \cite{susaki2014whole,stirman2016wide,fan2019video}, but particularly challenging when imaging non-flat living samples such as the human retina. Indeed, high-resolution is normally achieved with adaptive optics (AO) and scanning methods, which considerably reduce the useful FOV and increase the system complexity. An alternative technique is time-domain Full-Field Optical Coherence Tomography (FF-OCT) \cite{Dubois_04,Thouvenin_17}, which has already shown its potential for \textit{in-vivo} high-resolution retinal imaging \cite{mece2020high}. Here, we introduce curved-field FF-OCT, capable of matching the coherence gate geometry to the sample curvature, thus achieving a larger FOV than previously possible. Using this instrument, we obtained high-resolution images of living human photoreceptors close to the foveal center without AO and with a $1~mm~\times~1~mm$ FOV in a single shot. This novel advance enables the extraction of photoreceptor-based biomarkers with ease and spatiotemporal monitoring of individual photoreceptors. We compare our findings with AO-assisted ophthalmoscopes, highlighting the potential of FF-OCT, as a relatively simple, compact and low-cost system, to become a routine clinical imaging technique.


\end{abstract}

\vspace{1cm}


Owing to the optical properties of the eye, the retina is the only part of the central nervous system that can be visualized non-invasively \textit{in-vivo}, a crucial aspect for studying neuronal activity \cite{london2013retina,hunter2019imaging}. Due to their capacity to correct for static and dynamic ocular aberrations \cite{jarosz2017high,mece2019higher}, AO ophthalmoscopes have become the primary technique to image individual retinal neurons such as cone and rod photoreceptors in the living human retina \cite{liang1997supernormal,roorda2002adaptive, zacharria2011biomedical,gofas2018high}. Imaging individual retinal neurons \textit{in-vivo} with AO ophthalmoscopes has enabled new insights into retinal function such as colour vision \cite{zhang2019cone} and a better understanding of progression of retinal diseases such as age-related macular degeneration \cite{paques2018adaptive}. However, AO systems require quite complex, expensive and cumbersome hardware, strongly limiting their clinical and commercial deployment \cite{jonnal2016review,shemonski2015computational}. Thus, achieving high-cellular resolution in the living human retina without using AO is of tremendous interest.

Recent studies have achieved this goal by implementing a super-resolution optical-reassignment technique \cite{dubose2019super} or computational ocular aberration correction \cite{shemonski2015computational,hillmann2016aberration}. Nevertheless, they were not able to provide a real-time view of the retina and presented results in far foveal eccentricities (3$^o$ or more), where photoreceptors are most easily resolved \cite{shemonski2015computational,hillmann2016aberration}, or they used scanning illumination/detection techniques \cite{shemonski2015computational, dubose2019super}, which present an inherent low frame rate, limited FOV (lower than $0.3~mm \times 0.3~mm $) and high sensitivity to fixational eye motion \cite{mece2018fixational} leading to intraframe distortion. All these drawbacks affect the repeatability and accuracy of the measurement of essential biomarkers, such as photoreceptor distribution, density, and spacing, and the spatiotemporal monitoring of individual photoreceptors, all useful tools to diagnose retinal disorders at the earliest stage \cite{litts2017photoreceptor}.


An alternative modality to achieve high-cellular resolution without using AO is FF-OCT, which uses a spatially incoherent light source, a high-speed megapixel camera and time-domain phase modulation to acquire \textit{en-face} sections of the sample at a given depth. One attractive point of this technique for retinal imaging is the fact that the optical resolution of FF-OCT has a low sensitivity to predominant ocular aberrations (\textit{i.e.} defocus and astigmatism) \cite{Thouvenin_17,Xiao_16_osa,mece2019towards}. This interesting feature was recently highlighted in \cite{mece2020high}, where \textit{in-vivo} human cone mosaic at eccentricities close to the fovea was resolved. Although FF-OCT's FOV is theoretically only limited by the spatial sampling of the imaging camera, retinal curvature causes current images to have a limited useful FOV of about $0.4~mm \times 0.4~mm$ \cite{mece2020high, xiao2018vivo}.

Typically, FF-OCT uses a Linnik interferometer, where identical microscope objectives are placed in both reference and sample arms in a symmetric optical path configuration (Fig.~\ref{fig:SetUp}). When applied to retinal imaging, the FF-OCT symmetry is broken, as the sample arm no longer contains a microscope objective (Fig.~\ref{fig:SetUp}), but rather contains the anterior segment optics of the human eye. In this Letter, we experimentally demonstrate that when the symmetry is broken a curved coherence gate is generated, provoking a mismatch with the retinal curvature, and consequently limiting the useful FOV. We present novel advances in FF-OCT allowing for optical shaping of the geometry of the coherence gate, adapting it to the retinal geometry (Supplementary~Section~1,2), enabling the generation of single-shot, high-resolution, large FOV images ($1~mm \times 1~mm$) of the photoreceptor mosaic as close as 0.5$^o$ from the foveal center. Then, we show that important photoreceptor-based biomarkers together with the spatiotemporal monitoring of individual photoreceptors can be achieved without the need for AO in a relatively simple and low-cost optical imaging system.


Figure~\ref{fig:Target} shows how the coherence gate geometry can be optically manipulated by adding optical windows in one of the interferometer arms. When imaging the USAF target (Fig.~\ref{fig:Target}(a)) in a symmetric configuration the cross-section looks completely flat. The addition of an optical window in one of the arms breaks this symmetry, consequently, the flat sample presents an apparent curved cross-section. This apparent curvature happens because of a non-constant optical path difference along the FOV, generating a curved coherence gate (see Supplementary~Section~1). One can notice that the curvature increases as a function of the thickness of the optical window introduced in one of the arms. Depending on which arm the optical window is introduced in, the coherence gate curvature can present a concave or a convex shape. The induced curvature of the coherence gate directly impacts the useful FOV of an FF-OCT system, since it acquires \textit{en-face} curved sections of a flat sample at a given depth (Fig.~\ref{fig:Target}(c,d)). The same experiment can be done now using an OEMI-7 model eye (Ocular Instruments, Inc) in an asymmetric configuration (Fig.~\ref{fig:Target}(b)). In this case, already when no optical window is introduced, the coherence gate presents a significant curvature. The introduction of an optical window in the sample arm enables compensation for the initial apparent curvature, increasing the useful FOV (Fig.~\ref{fig:Target}(e,f)).

\begin{figure}[!ht]
\centering
\includegraphics[width=\linewidth]{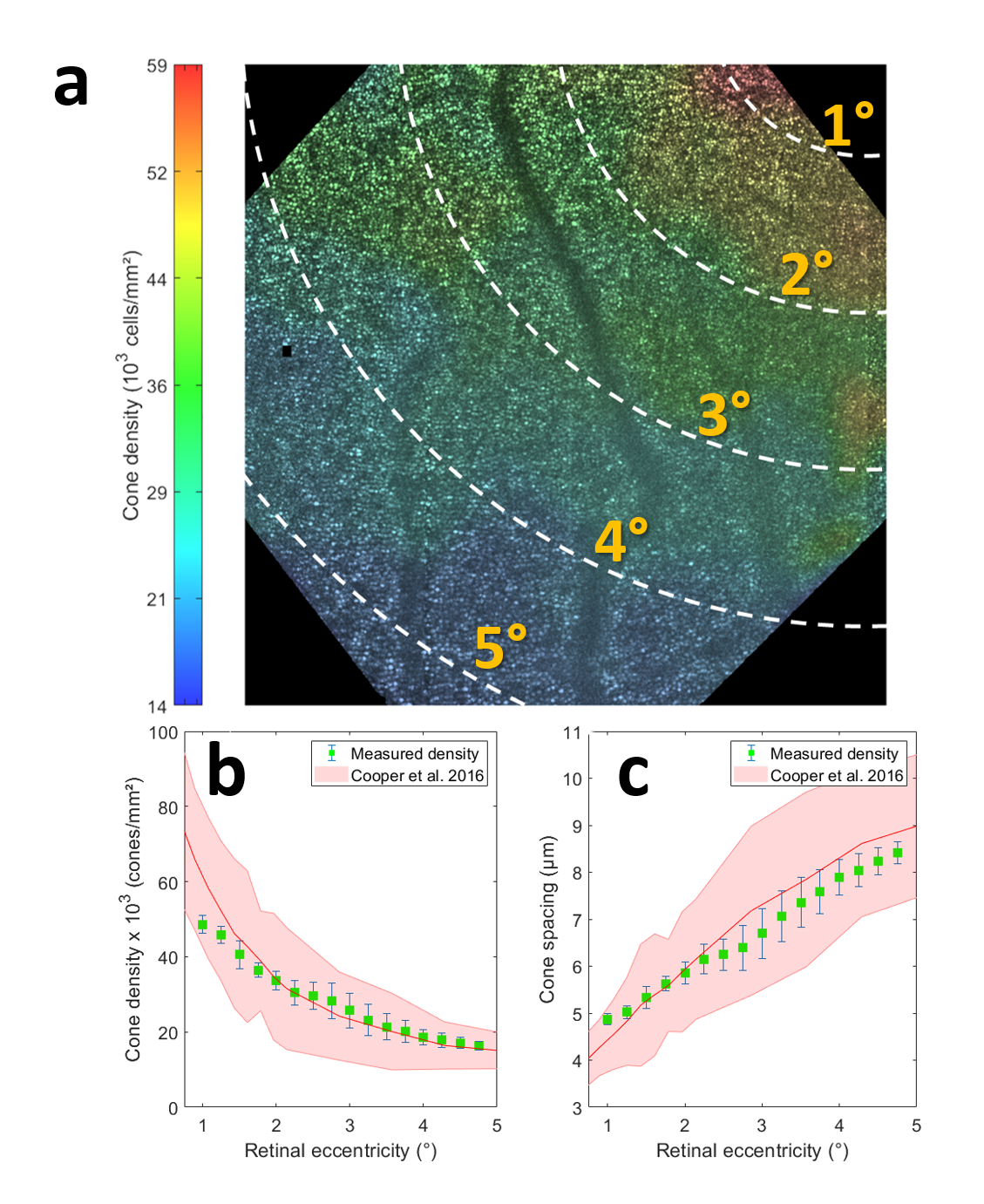}
\caption{Measurement of photoreceptor-based biomarkers. \textbf{a,} Cone density distribution color map. \textbf{b,c,} Respectively, the estimated cone density and spacing mean (green squares) and standard deviation (blue lines) values as a function of the retinal eccentricity computed from \textbf{a}. The red line and shaded region are respectively the mean computed density and the 95\% prediction interval obtained from a dataset of a 20-eye healthy population using AO scanning-laser ophthalmoscope \cite{cooper2016evaluating}. }
\label{fig:Density}
\end{figure}

\textit{In-vivo} retinal imaging from the inner/outer segment junction (IS/OS) was performed in a healthy volunteer before and after compensation of the coherence gate curvature in an area as close as 0.5$^o$ from the foveal center (Fig.~\ref{fig:InVivo}). The addition of an N-BK7 optical window with $20~mm$ thickness enabled achievement of high-resolution foveal cone mosaic imaging with a useful FOV of $1~mm \times 1~mm$, larger than the previous FOV by a factor of 6.25. Zoomed areas and their respective power spectral density (PSD) are also presented. All PSDs presented Yellot's ring, the spectral signature of the photoreceptor mosaic image \cite{yellott1982spectral}. The high-resolution allied with the large FOV enabled measurement of photoreceptor-based biomarkers within a few seconds on a single-shot (no image montaging is necessary) with ease (Fig.~\ref{fig:Density}(a)). Figures \ref{fig:Density}(b,c) show how the cone density and spacing respectively evolve with eccentricity. All measurements are consistent with histology and AO-assisted cone density measurement \cite{curcio1990human,cooper2016evaluating}.

Minimizing the effect of the coherence gate curvature also helps to increase the repeatability and consistency of the imaging, allowing for spatiotemporal monitoring of individual photoreceptors. Supplementary~Video~5  presents four image sequences acquired in the same region with a $10~min$ interval between acquisitions. Note in zoomed areas that individual photoreceptors can be tracked frame-by-frame ($3~ms$ interval) and video-by-video ($10~min$ interval). Green boxes highlight cones that are visible in all four videos, while yellow boxes those visible for some of the videos, which might indicate a change of reflectivity due to phototransduction \cite{litts2017photoreceptor} (Supplementary~Section~4). Red hexagons highlight the typical pattern found in cone mosaic close to the fovea \cite{curcio1990human,cooper2016evaluating}.  Figure~\ref{fig:Monitoring}(a-d) presents the averaged image obtained from zoomed areas of Supplementary~Video~5. Figures \ref{fig:Monitoring}(e-h) and Supplementary~Videos~6,7 show a cone-to-cone comparison of FF-OCT image with images generated by two AO ophthalmoscope modalities for the same subject and observed region. FF-OCT allows visualization of most, if not all, photoreceptors identified in AO-assisted images. Note in Fig.~\ref{fig:Monitoring}(i) that PSDs from all FF-OCT images present a clear and well-defined cone mosaic spatial frequency (red-area) at 120 cycles/mm, as it is the case for both AO-assisted imaging modalities. On the other hand, when using a scanning laser ophthalmoscope without closing the AO-loop, an erroneous spatial frequency is measured, leading to an error in cone density computation.

Through the optical manipulation of the coherence gate geometry, the data shown here represent a clear improvement in retinal imaging, allying high cellular-resolution in all three dimensions, large FOV, and high-acquisition rate of a given retinal plane (300 Hz), without using any optical or digital aberration compensation other than prescribed eyeglasses, considerably simplifying the hardware and software complexity. This novel performance provides four-dimensional (4D) monitoring of individual photoreceptors and computation of photoreceptor-based biomarkers with good accuracy and repeatability, preventing image distortion and long, fastidious acquisition sessions to cover the retinal area of interest, as is the case with scanning systems. 


In conclusion, we have developed curved-field FF-OCT, an imaging system capable of adapting its coherence gate geometry to the curvature of the sample being imaged, making large FOV imaging possible while keeping high-resolution in samples under constant motion, such as the living human eye. While this paper discusses the application of curved-field FF-OCT in retinal imaging, it can be extended to other non-flat samples, such as the cornea \cite{Mazlin_18} and retinal organoids \cite{scholler2019motion}. Moreover, other optical elements than optical windows could also be used to manipulate the coherence gate shape (Supplementary Section 3). We have shown both qualitative and quantitative improvement in retinal imaging allowing FF-OCT to monitor individual photoreceptors in 4D with ease and to quickly and robustly compute photoreceptor-based biomarkers close to the foveal center without using AO. The relative simplicity and low-cost of the curved-field FF-OCT imaging technique compared to current AO-assisted ophthalmoscopes may pave the way towards the adoption of curved-field FF-OCT as a routine clinical imaging system, providing a new understanding of retinal structure, function, and diseases.

\section*{Acknowledgements}

This work was supported by grants entitled “HELMHOLTZ” (European Research Council (ERC) \#610110).

\section*{Author contributions}

P.M. designed and constructed the optical system, analysed the results and drafted the manuscript. P.M. and C.B. developed the presented method. P.M., K.Gro and K.Gri. collected data. P.M. and J.S. wrote the software for the imaging system and for data processing. P.M., K.Grou. J.S., O.T. and C.B. discussed the results. C.B., K. Gri. and M.F. provided overall guidance to the project. M.F. obtained funding to support this research. All authors reviewed and edited the manuscript.

\section*{Competing interests}
P.M. and C.B. are listed as inventors on a patent application (application no. 19306683.4) related to the work presented in this manuscript. All other authors have nothing to disclose.

\newpage
\begin{figure*}[!hb]
\centering
\includegraphics[width=\linewidth]{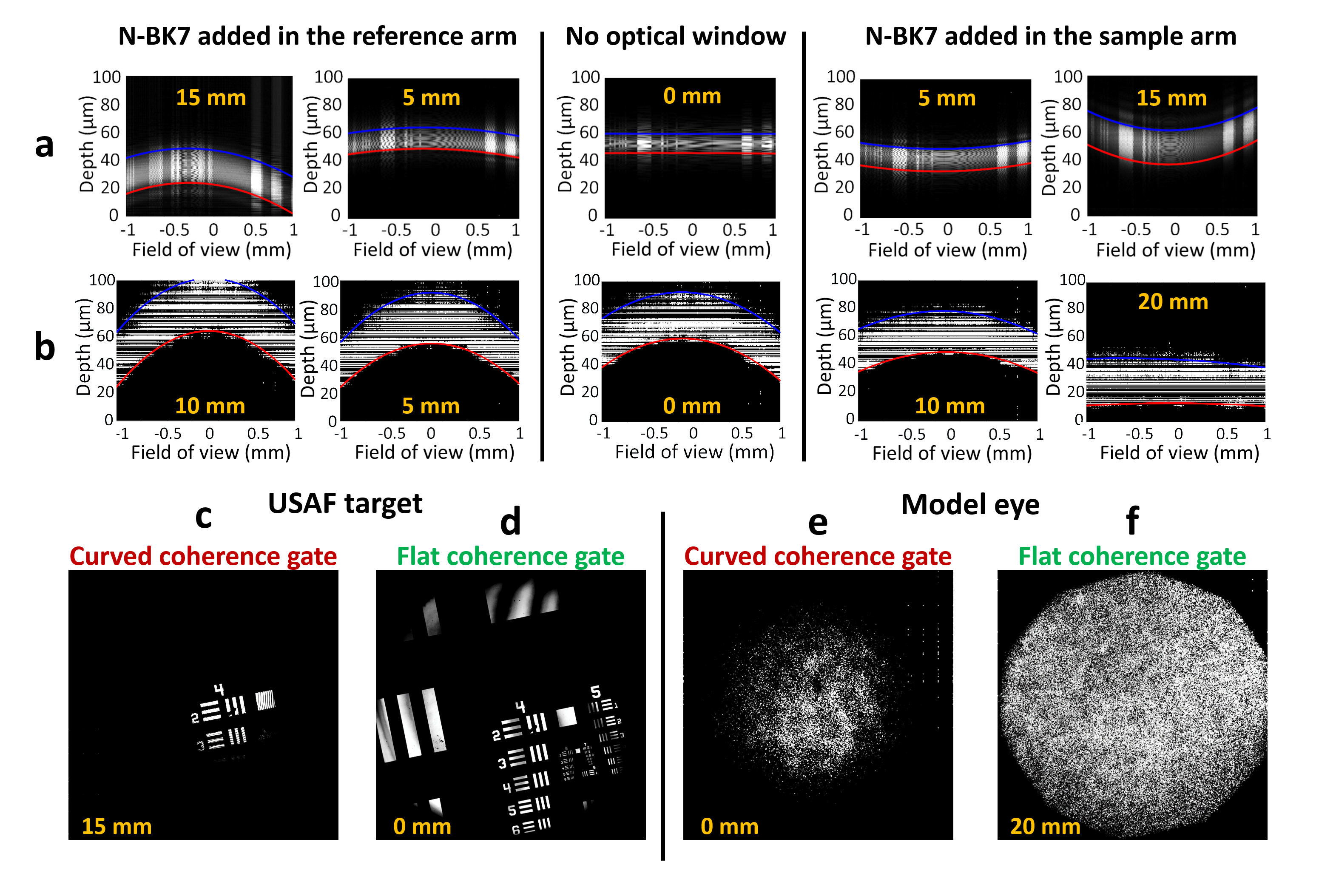}
\caption{Optical manipulation of the coherence gate geometry and influence on the FOV. \textbf{a-b,} Cross-sections after introducing N-BK7 optical windows with various thicknesses (in yellow) in the reference or sample arms when imaging, respectively, the USAF target and the model eye. Note for the model eye imaging, that the curvature of the coherence gate is quite important when no optical window is introduced: the difference between the center and the edge is around 20 $\mu m$. Measured and theoretical optical path differences as a function of the FOV for different optical window thicknesses, as well as their respective degree of curvature, are given in Supplementary~Figs.~\ref{fig:Theory}(a-d). \textbf{c-f,} The effect of the coherence gate geometry over the FF-OCT useful FOV. When curvature is compensated, the FOV is increased by a factor 6.25. Supplementary~Videos~1-4 show acquired Z-stacks for curved and flat coherence gate when imaging the USAF target and the model eye.}
\label{fig:Target}
\end{figure*}

\newpage

\newpage

\begin{figure*}[!ht]
\centering
\includegraphics[width=\linewidth]{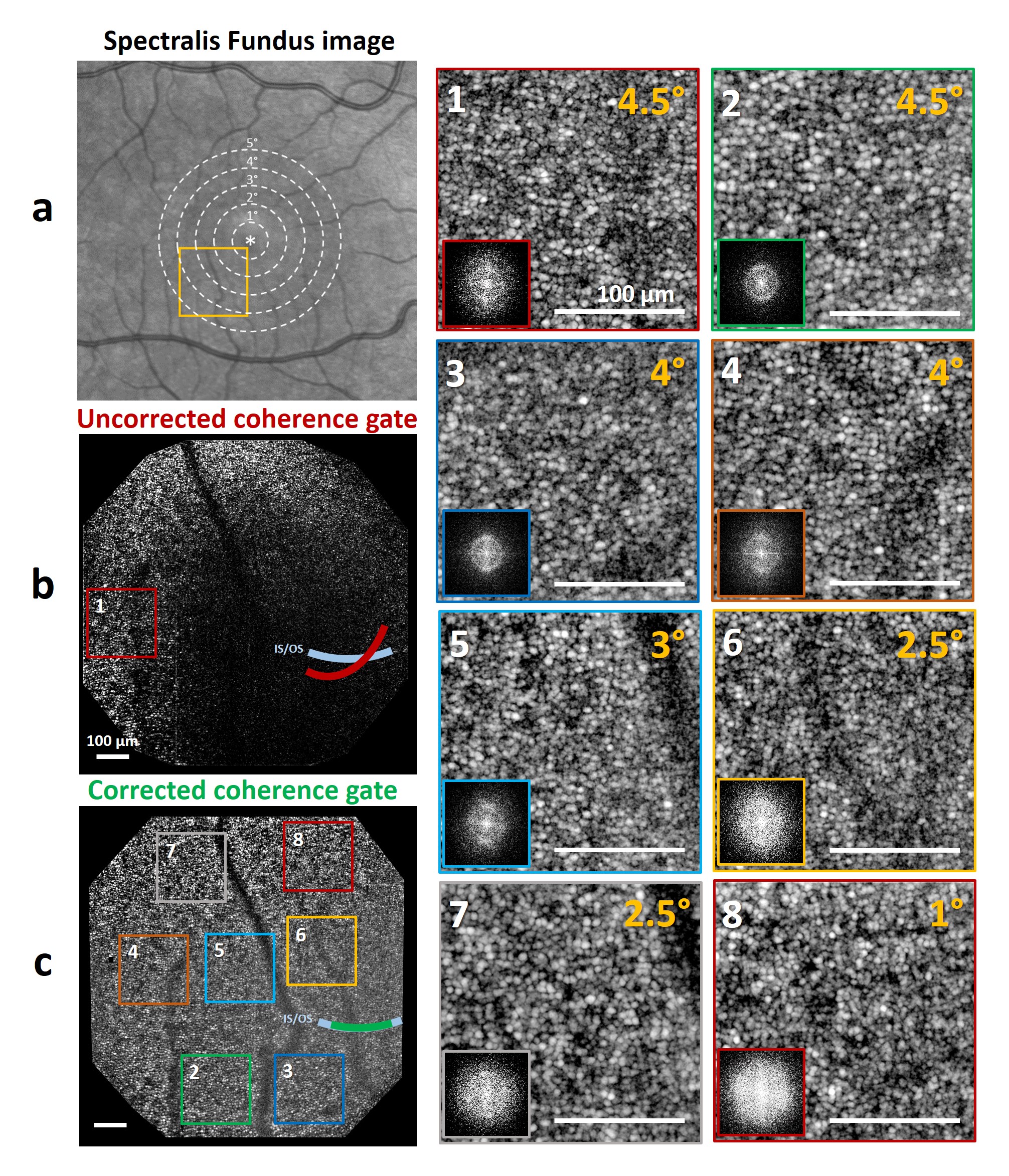}
\caption{\textit{In-vivo} retinal imaging allying large FOV and high-resolution. \textbf{a,} Retinal fundus image obtained using Spectralis scanning-laser ophthalmoscope, where the yellow square indicates the location of acquired images, \textit{i.e.} as close as $0.5^o$ from the foveal center. \textbf{b,c,} Respectively, acquired images from the IS/OS junction before and after correcting for the coherence gate curvature. Note that, with a curved coherence gate, besides reduced useful FOV, the system becomes very sensitive to the 3D alignment of the subject in front of the system \cite{mece2019towards}. This effect is provoked by the non-coincidence of the curvature center and the FOV center. Subject alignment is better managed when the strong coherence gate curvature is compensated. Zoomed areas, and their respective eccentricity (in yellow) and PSD are also presented. All PSDs present the Yellot's ring \cite{yellott1982spectral}. Scale bar, $100 \mu m$.}
\label{fig:InVivo}
\end{figure*}

\newpage

\newpage

\begin{figure}[!ht]
\centering
\includegraphics[width=\linewidth]{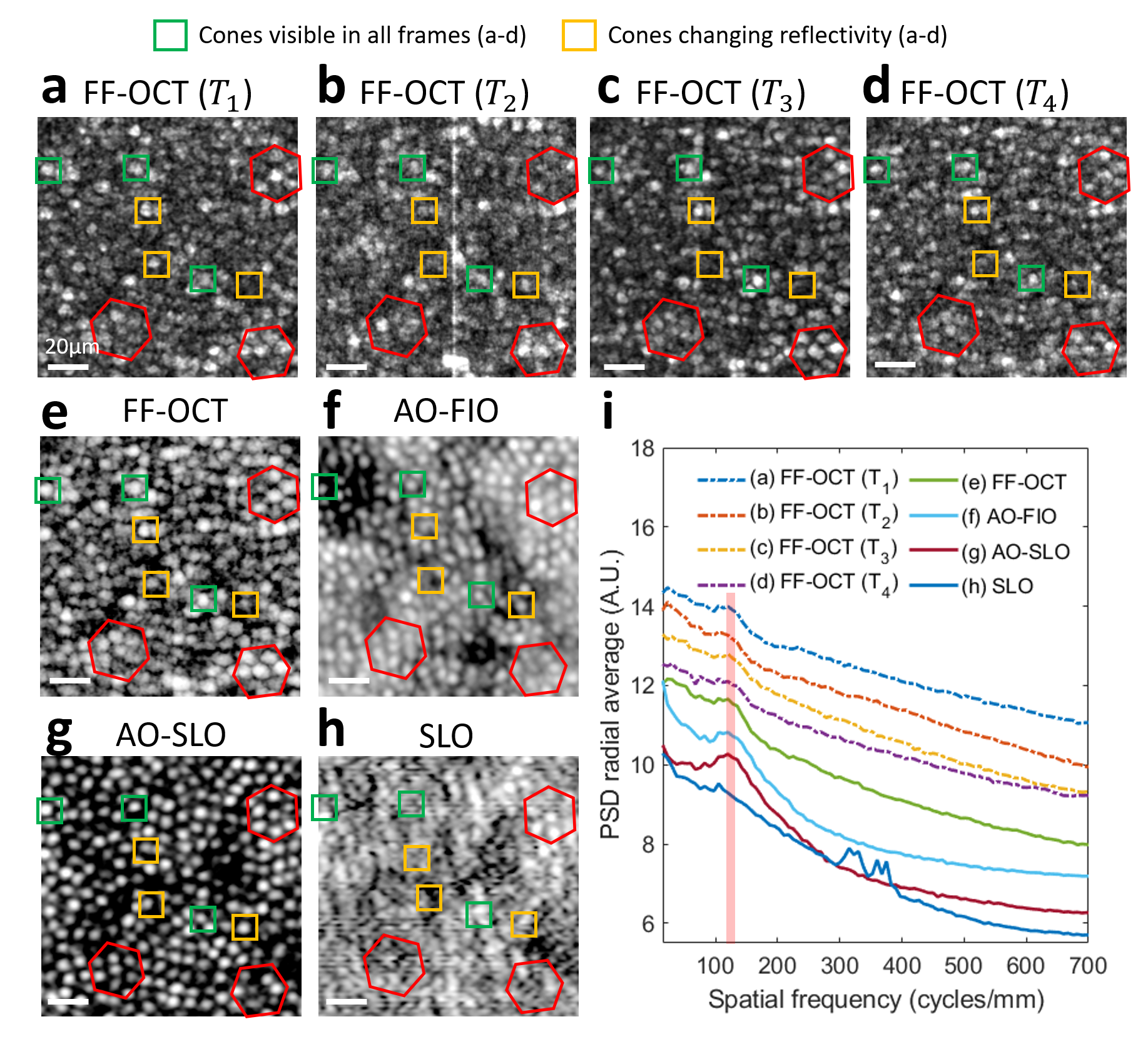}
\caption{4D of individual photoreceptors and comparisons with AO-assisted ophthalmoscopes. \textbf{a-d,} Cone mosaic images acquired at the same region ($4.5^o$ eccentricity) for different time points $T_{1,2,3,4}$, with 10 minutes interval between acquisitions, highlighting the capacity of curved-field FF-OCT to track individual cones in space and time with ease. Green boxes: cones visible for all four acquisitions. Yellow boxes: cones changing reflectivity. Red hexagons: pattern found in cone mosaic close to the fovea \cite{curcio1990human}. \textbf{e,} Cone mosaic image obtained after averaging images from \textbf{a-d}. \textbf{f-h,} Respectively, cone mosaic image at the same region acquired using AO-ophthalmoscopes. FIO: Flood-illumination ophthalmoscope. SLO: confocal scanning-laser ophthalmoscope. \textbf{h,} Acquisition made without closing the AO-loop. Supplementary~Videos~6,7 show a cone-to-cone comparison between FF-OCT image and, respectively, AO-FIO and AO-SLO. \textbf{i,} PSD radial average for images \textbf{a-h}, where the red-area outlines the expected cone mosaic spatial frequency. PSDs were verticaly displaced for better visualization. Scale bar, 20 $\mu m$.}
\label{fig:Monitoring}
\end{figure}

\newpage
\section*{Methods}
\subsection*{Samples imaged}
To investigate the impact of the asymmetric Linnik interferometer configuration on coherence gate curvature, three samples were imaged: 1) a standard USAF resolution target; 2) an OEMI-7 model eye (Ocular Instruments, Inc) which takes into account the eye geometry, optical power and dispersion properties (henceforth, named model eye); and 3) the retina of a healthy volunteer presenting a refractive error of $1.5~D\times 0.5~D\times 160^o$ (spherical $\times$ cylindrical $\times$ axis of cylindrical errors). Research procedures followed the tenets of the Declaration of Helsinki. Informed consent was obtained from the subject after the nature and possible outcomes of the study were explained. This study was authorized by the appropriate ethics review boards (CPP and ANSM (IDRCB number: 2019-A00942-55)).

\subsection*{Experimental setup}
\begin{figure}[!ht]
\centering
\includegraphics[width=0.7\linewidth]{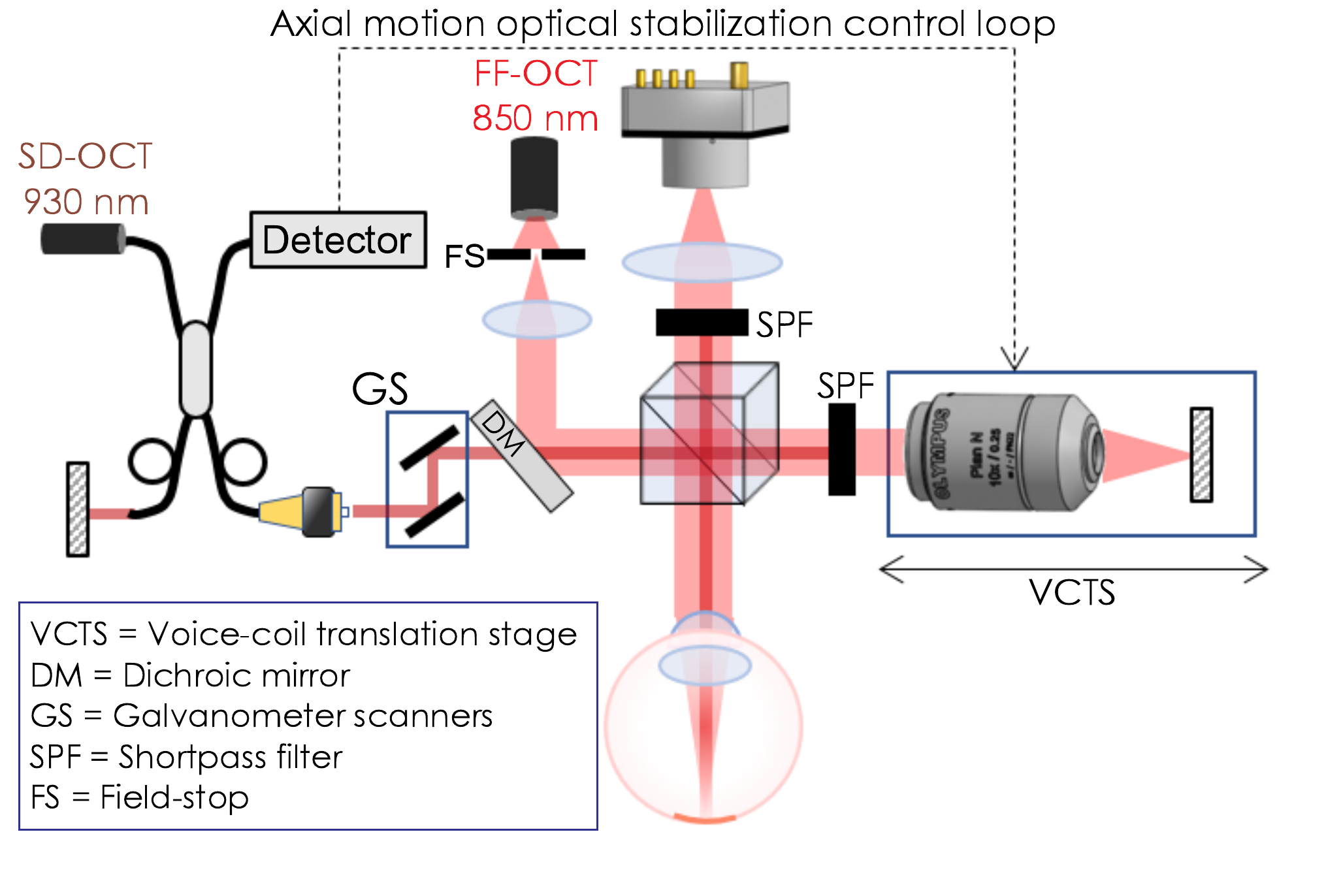}
\caption{Schematic drawing of the custom-built FF-OCT system coupled with an SD-OCT for real-time axial motion correction and FF-OCT coherence gate positioning guidance. In the case of the USAF target imaging, a microscope objective, identical to the one used in the reference arm, is used leading to a symmetric Linnik configuration (not shown in the schematic). In the case of the model eye and \textit{in-vivo} retinal imaging, no microscope objective is placed in the sample arm, leading to an asymmetric Linnik configuration (shown in the schematic).}
\label{fig:SetUp}
\end{figure}

Figure~\ref{fig:SetUp} presents the schematic of the custom-built FF-OCT system coupled through a dichroic mirror with a Thorlabs Ganymede-II SD-OCT system. The FF-OCT comprises a light-emitting diode (LED) with $\lambda$ = 850 nm center wavelength and 30 nm bandwidth (M850L3, Thorlabs), used as a spatially incoherent illumination source, giving a theoretical axial resolution of approximately 8 $\mu$m in water. The LED is focused by a condenser $20~mm$ in front of the eye's pupil. A physical diaphragm is positioned in front of the LED, conjugated to the retina and the FF-OCT reference mirror. The illumination beam is split into the reference and the sample arms by a 50:50 cubic beam splitter (BS). For the reference arm, an Olympus 10X/0.25 NA Plan Achromat objective is used with a silicon mirror placed at the focal plane of the objective. The whole reference arm (microscope objective and silicon mirror) is mounted on a fast voice-coil translation stage (X-DMQ12P-DE52, Zaber Technologies Inc.), allowing for adjustment of the coherence gate position. For the sample arm, two configurations were used. In the case of the USAF target imaging, a microscope objective, identical to the one used in the reference arm, is used leading to a symmetric Linnik configuration (not shown in Fig.~\ref{fig:SetUp}). In the case of the eye model and \textit{in-vivo} retinal imaging, both were aligned along the optical axis, without the use of a microscope objective leading to an asymmetric Linnik configuration (shown in Fig.~\ref{fig:SetUp}). The FF-OCT light beam arrives with an 8 mm diameter in the eye's pupil. The back-scattered photons from both arms are recombined by the same BS and focused onto a high-speed (up to 720 Hz at full-frame) CMOS camera (Q-2A750-Hm/CXP-6, Adimec) for FF-OCT imaging. In the case of \textit{in-vivo} retinal imaging, the SD-OCT and the voice-coil translation stage were used to measure and correct for involuntary axial eye movements during the imaging acquisition in a closed-loop fashion \cite{mece2020high}.

\subsection*{Image acquisition}
In the case of the USAF target and the model eye imaging, a total of 200 images from different depths were acquired at 10 Hz, by moving the voice-coil translation stage of the reference arm at a constant speed of 5 $\mu$m/s. This procedure allowed us to acquire \textit{en-face} images from different depths, forming Z-stacks. This procedure was repeated after adding optical windows of N-BK7 of different thicknesses in the reference or sample arms in order to change the geometry of the coherence gate. 

In the case of living human retinal imaging, The subject was seated in front of the system and stabilized with a chin and forehead rest and asked to fixate a target placed at an infinite focal conjugate. During the imaging session, the subject was wearing her prescription glasses to increase the signal level of FF-OCT \cite{mece2019towards,mece2019real}. Image acquisition was realized in a dark room, maximizing the pupil dilation. Phase modulation was performed by the residual axial motion after optical stabilization \cite{mece2020high}. Image sequences were composed of 150 images acquired at 300 Hz using custom-built software \cite{FFOCT_JS}. The FF-OCT camera worked close to saturation to use the whole full well capacity, decreasing relative importance of shot noise \cite{scholler2019probing}. During image acquisition, the total power entering the eye from the FF-OCT illumination source and the SD-OCT scanning source were respectively 1.3~mW (for 0.5~s) and 0.25~mW (continuous scanning), which are below the ocular safety limits established by the ISO standards for group 1 devices.

\subsection*{Image processing}
\label{Model}
In the case of the USAF target and the Ocular model eye imaging, each Z-stack was digitally converted into axial sections, \textit{i.e.} a cross-sectional view of the sample. This step was possible by taking into account only the pixels of a central row of each image. Next, image segmentation was applied using an intensity-based thresholding algorithm and a least square parabola fitting algorithm, enabling to measure the degree of curvature of converted cross-sections.

For \textit{in-vivo} retinal imaging, each image was normalized by dividing itself by its mean value. Since the phase was randomly modulated by the residual tracking error, and to eliminate the incoherent terms, we adopted a 2-phase demodulation \cite{mece2020high,Scholler_19}. The 2-phase demodulation consists of subtracting one image $I_{N}$ from the next $I_{N+1}$ and taking the modulus. Next, images with a very low or absent useful signal, mainly due to an insufficient phase shift between consecutive images,  were automatically detected using an intensity-based threshold algorithm, and then excluded from the image sequence. Finally, useful images were registered using a custom-built normalized cross-correlation algorithm, where the image presenting the highest signal level was chosen as the reference.

\subsection*{Cone density map}
To generate the pointwise density map, we divided the cone mosaic image into an overlapping grid of $300\times 300$ pixels (corresponding to $200~\mu m \times 200~\mu m$) regions of interest (ROIs), where each ROI was displaced from the previous by $30$ pixels (corresponding to $20 \mu m$). These values were chosen empirically to provide a good trade-off between pointwise accuracy and map smoothness. Then, cone density and spacing were computed for each ROI using a fully automated algorithm based on modal spacing as described in \cite{cooper2019fully}. Image interpolation using splines was used to increase the size of the cone density map in order to match the cone mosaic image. 

\subsection*{Other used imaging modalities}
Retinal images of the same subject were also acquired using the following instruments: Spectralis scanning-laser ophthalmoscope (Heidelberg Engineering, Germany) - Fig.~\ref{fig:InVivo}(a); rtx-1 adaptive optics flood-illumination ophthalmoscope (Imagine Eyes, France) - Fig.~\ref{fig:Monitoring}(f) and Supplementary~Video~6;  MAORI confocal adaptive optics scanning-laser ophthalmoscope (Physical Sciences, Inc, Andover, MA, USA) - Fig.~\ref{fig:Monitoring}(g,h) and Supplementary~Video~7. All images were acquired in the same conditions as previously described: image acquisition was realized in a dark room, maximizing the pupil dilation and the subject was wearing her prescription glasses.

\subsection*{Data availability}
The study data are available from the corresponding author upon request.

\subsection*{Code availability}
The acquisition software is available at \cite{FFOCT_JS}.
\newpage
\bibliographystyle{ieeetr}
\bibliography{CurvedFFOCT}

\newpage
\section{Supplementary information}
\paragraph{Video 1} Z-stack of the USAF target for a curved coherence gate, after adding a 15 mm thick N-BK7 optical window in the sample arm and breaking the symmetric configuration of the interferometer.

\paragraph{Video 2} Z-stack of the USAF target for a flat coherence gate, \textit{i.e.} when no optical window is introduced in the inteferometer, in a symmetric configuration of the interferometer. 

\paragraph{Video 3} Z-stack of the model eye when no optical window is introduced in the inteferometer. The asymmetry of the reference and the sample arms provoke the appearance of a curved coherence gate.

\paragraph{Video 4} Z-stack of the model eye after adding a 20 mm thick N-BK7 optical window in the sample arm, correcting for the coherence gate curvature.

\paragraph{Video 5} Four image sequences acquired in the same region for the same subject, with a 10 minute interval between acquisitions, after compensating for coherence gate curvature. Note in the zoomed areas, that individual photoreceptors can be monitored frame-by-frame (3 ms interval) and video-by-video (10 minutes interval). Green boxes highlight cones that are visible in all four videos. Yellow boxes highlight cones that present a temporal variability in reflectivity, probably due to phototransduction \cite{jonnal2007vivo, litts2017photoreceptor}. Red hexagons highlight the typical pattern in cone mosaic close to the fovea \cite{curcio1990human, cooper2016evaluating}.

\paragraph{Video 6} Cone-to-cone comparison between curved-field FF-OCT cone mosaic image and the rtx-1 AO flood-illumination ophthalmoscope image for the same subject at the same region.

\paragraph{Video 7} Cone-to-cone comparison between curved-field FF-OCT cone mosaic image and the PSI MAORI
confocal AO scanning-laser ophthalmoscope image, with and without closing the AO-loop, for the same subject at the same region.

\subsection{Curved coherence gate caused by full-field interferometer asymmetry}

The optical path length (OPL) when light propagates in a material $x$ of thickness $e_x$, refractive index $n_x$ for a given angle $\theta_x$ can be expressed as:
\begin{equation}
    OPL(\lambda) = \sum_{x = 1}^N n_x(\lambda) \frac{e_x}{cos\theta_x}
    \label{Eq:Model}
\end{equation}

Where $n_x(\lambda)$ is given by the Sellmeier equation \cite{sellmeier1871erkarung}. The incident angle in a material $x$, $\theta_x$, can be expressed as a function of the incident angle in the air $\theta_1$, as follows:
\begin{equation}
   \theta_x = sin^{-1}\left(\frac{sin\theta_1}{n_x(\lambda)}\right)
    \label{Eq:theta}
\end{equation}

Through Eqs. \ref{Eq:Model} and \ref{Eq:theta}, one can notice the dependency of the OPL on the wavelength ($\lambda$) and the incident angle in the air ($\theta_1$):

\begin{equation}
    OPL(\lambda, \theta_1) = \sum_{x = 1}^N n_x(\lambda) \frac{e_x}{cos\Big(sin^{-1}\left(\frac{sin\theta_1}{n_x(\lambda)}\right)\Big)}
    \label{Eq:FinalModel}
\end{equation}

The depth position of the \textit{en-face} imaged generated by the FF-OCT, \textit{i.e.} the coherence gate, is given by the the optical path difference (OPD) between the reference and sample arms: 

\begin{equation}
    OPD (\lambda, \theta_1) = OPL_{reference}(\lambda, \theta_1) - OPL_{sample}(\lambda, \theta_1)
    \label{Eq:OPD}
\end{equation}

If the OPL of both arms are symmetric, as it is the case when two microscope objectives are placed in both arms, the OPD is null for all values of $\theta_1$, \textit{i.e.} along the FOV. On the other hand, an asymmetry of both OPLs creates a dependency on the the wavelength ($\lambda$) and the incident angle in the air ($\theta_1$), \textit{i.e.} the FOV. In this last case, OPD would only be null for a given incident angle, and it would increase as this angle is increased, the origin of the curvature of the coherence gate.

\begin{figure}[!ht]
\centering
\includegraphics[width=\linewidth]{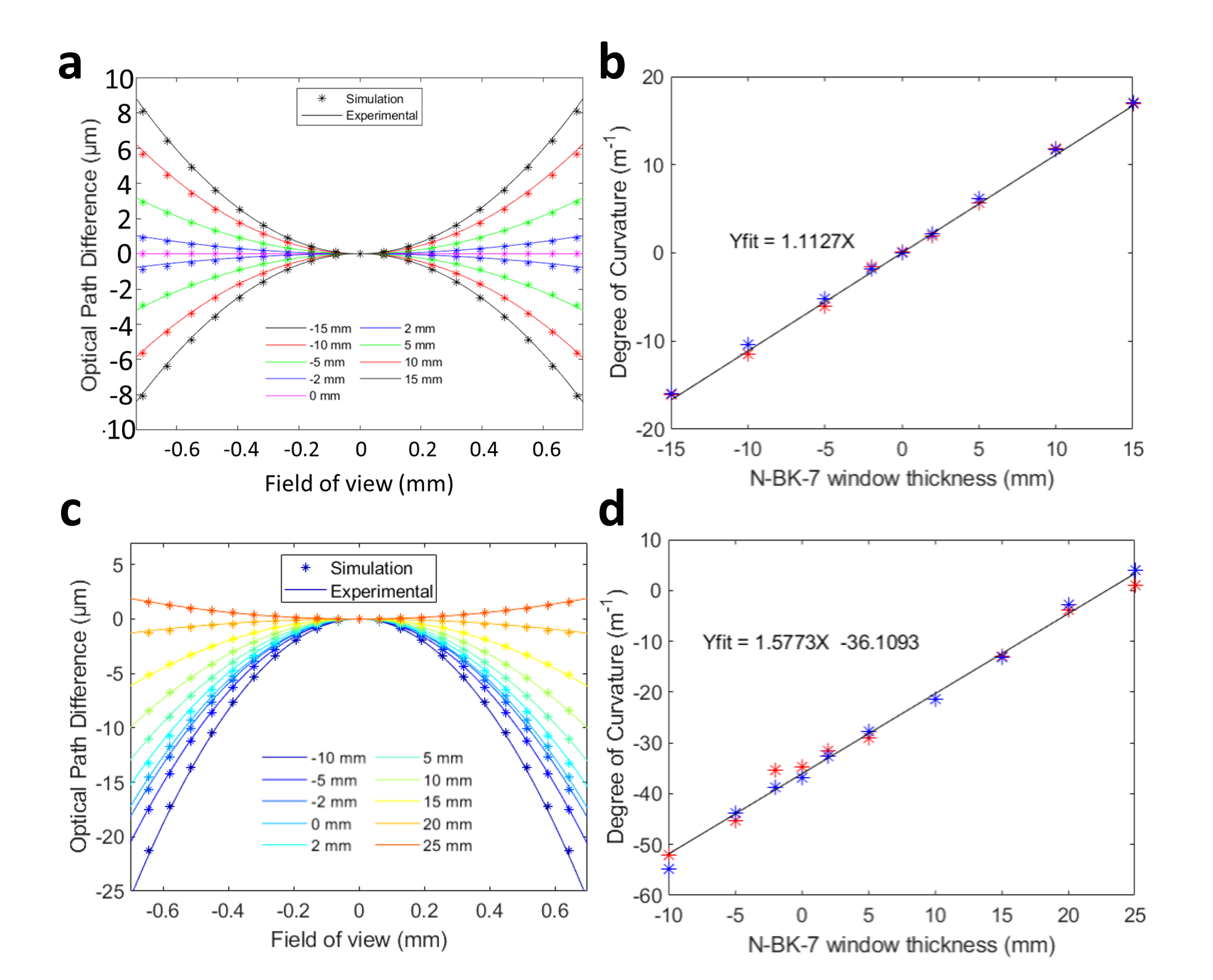}
\caption{Optical path difference along the FOV. (a,c) Experimental data and theoretical values of the optical path difference as a function of the FOV for the USAF target and the model eye imaging respectively. Different curvature plots present the effect on the coherence gate after introducing N-BK7 optical windows, of various thicknesses, in the reference arm (negative values) or in the sample arm (positive values). \textbf{(b,d)} Degree of curvature as a function of the N-BK7 optical window thickness, where zero degrees of curvature means a flat profile, for the USAF target and the model eye imaging respectively.}
\label{fig:Theory}
\end{figure}

Figue \ref{fig:Theory}(a-d) compare our findings in \ref{fig:Target}(a,b) with the expected theoretical optical path difference (OPD) for the USAF target and the model eye imaging, according to Eqs. \ref{Eq:FinalModel} and \ref{Eq:OPD}. 

The appearance of the coherence gate curvature was previously reported for conventional spectral-domain raster-scanning OCT, mainly linked to the scanning-induced path length variation in non telecentric optical systems \cite{podoleanu1998transversal, graf2010correction}. They also reported that this effect could be largely ignored in low-NA imaging systems \cite{graf2010correction}, as it is the case of the eye in retinal imaging application. In the present Letter, we show that the coherence gate curvature is amplified, and no longer negligible for low-NA FF-OCT, when the symmetry of the Linnik interferometry is broken.

Some solutions to avoid this deformation of the coherence gate, in high-NA OCT, are: the acquisition of images with stage scanning \cite{vinegoni2006integrated}, using a common path interferometer \cite{ralston2007interferometric}, or correcting it in a computationally manner by recovering the phase of the analytic signal from calibration data \cite{graf2010correction}. Nevertheless, either these approaches greatly limit the practical use of this technique, or cannot be applied for \textit{in-vivo} retinal imaging conditions, or it is not adapted for a FF-OCT modality, since all points in the field-of-view are collected in parallel manner and not in a sequential manner as in conventional OCT.

\subsection{Dispersion effect}
\begin{figure}[!ht]
\centering
\includegraphics[width=\linewidth]{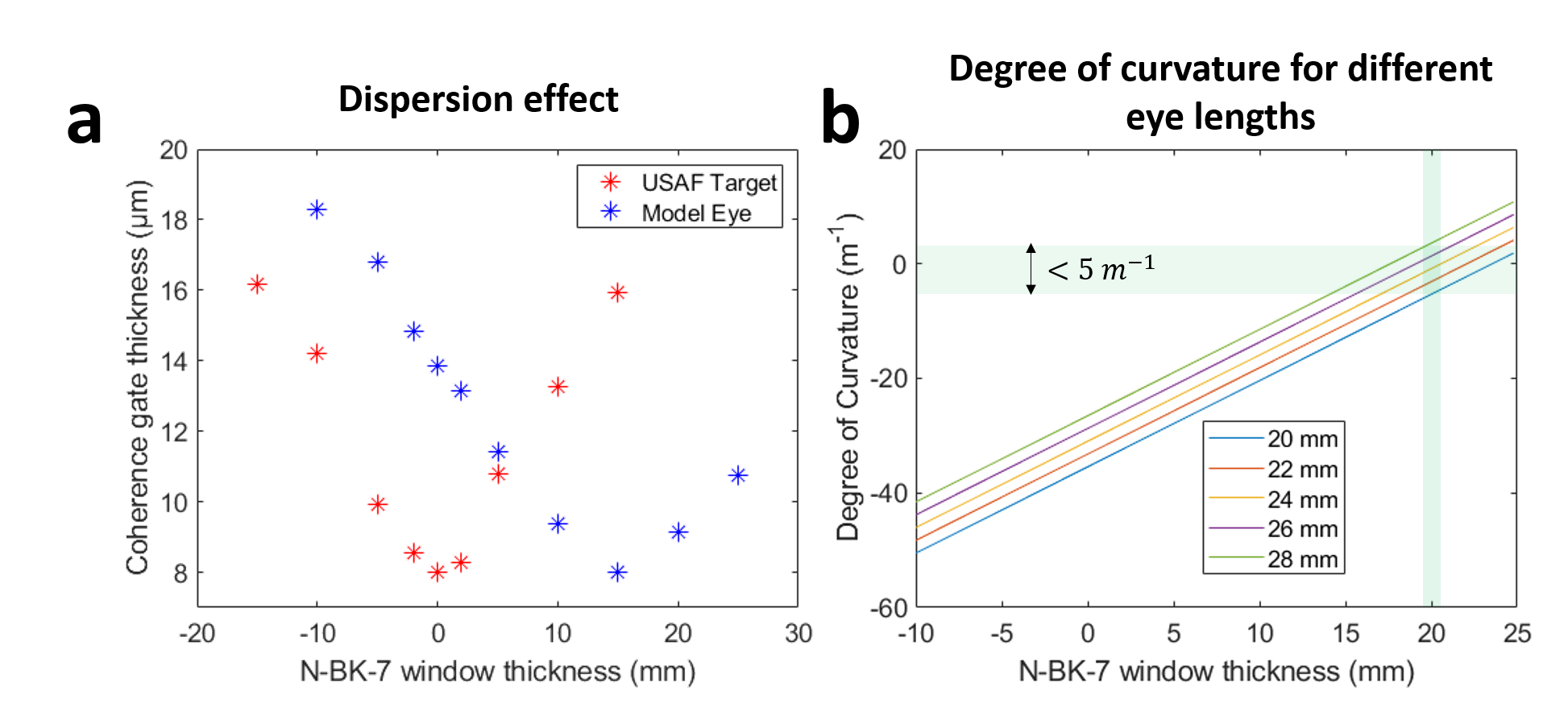}
\caption{Dispersion effect and eye-length. \textbf{(A)} FWHM of the coherence gate thickness (FF-OCT axial resolution) as a function of the optical window thickness in the case of the USAF target (red points) and model eye (blue points) imaging respectively. \textbf{(B)} Expected theoretical value of degree of curvature (computed using Eq. \ref{Eq:Model}) as a function of the thickness of the N-BK7 introduced to compensate for coherence gate curvature for different eye lengths. Green shaded area highlights that a N-BK7 of 20 mm thickness covers the majority of eye lengths in the population (from 20 mm to 28 mm) \cite{kolb2007gross}.}
\label{fig:Dispersion}
\end{figure}

Another expected effect brought by the asymmetry of the Linnik interferometry is dispersion, affecting the axial resolution of the FF-OCT (\textit{i.e.} axial sectioning capacity). This phenomenon can be noted in cross-sections of Figs. \ref{fig:Target}(a,b) as an increase of the coherence gate thickness (distance between the blue and red curves). Figure~\ref{fig:Dispersion} (a) shows the FWHM of the coherence gate thickness (FF-OCT axial resolution) as a function of the optical window thickness in the case of the USAF target imaging (in red) and the model eye imaging (in blue). Note that for the USAF target imaging, a flat coherence gate coincides with the best axial sectioning capability, both happening when no optical window is added (symmetric configuration). In the case of the model eye imaging, not surprisingly, the optimum correction of dispersion does not coincide with the optimum correction of coherence gate curvature, since these two phenomena have different physical origins (Supplementary Section 1). We saw that an optimum compensation of the coherence gate curvature would be possible with a 22 mm N-BK7 optical window. Fig.~\ref{fig:Dispersion} (a) shows, for the model eye imaging, that an optimum dispersion compensation would be possible with a 15 mm of N-BK7 optical window. In this study, in order to compromise and minimize the effects of both curvature of the coherence gate and dispersion for \textit{in-vivo} retinal imaging, we used a N-BK7 optical window with 20 mm thickness. According to Fig.~\ref{fig:Theory}(c) and Fig.~\ref{fig:Dispersion}(a), a good trade-off could be found in this case: the axial resolution would be around 9$\mu$m, instead of 8$\mu$m, and the degree of curvature would be lower than 5, meaning an edge-center difference of 1$\mu$m. Figure~\ref{fig:Dispersion}(b) shows that this correction would be suitable for eye lengths varying from 20 mm to 28 mm, covering the majority of cases in the population \cite{kolb2007gross}.

\subsection{Adapting the interferometer reference arm to the imaged sample}
In this study, we only considered N-BK7 optical windows to minimize effects of both coherence gate curvature and dispersion. Another solution to optically manipulate the geometry of the coherence gate is the introduction of a deformable reference mirror in the reference arm. This solution would decouple the correction of the coherence gate curvature and dispersion, since a suitable optical window in the sample arm could compensate for dispersion effect and the deformable reference arm would correct the residual curvature of the coherence gate, without affecting dispersion, improving signal level, axial resolution and possibly achieving even larger FOV. Moreover, this solution might extend the curved-field FF-OCT to samples with irregular surface, more complex than a curvature. Nevertheless, the main drawback of this solution is the considerable increase of the system complexity and cost, making clinical transfer challenging. A solution considering only an optical window, as initially presented, seems sufficient to be able to optimize for coherence gate geometry and dispersion for a large population (Supplementary Section 2).

\subsection{Photoreceptor reflectivity variation}
One of the most interesting features of photoreceptor mosaic is the temporal and the cone-to-cone variability in reflectivity \cite{pallikaris2003reflectance}. Figures \ref{fig:Monitoring} (a-d) and Supplementary Video 5 present both reflectivity change over time (seconds and minutes) and from cone to cone. This short-term intrinsic variability in reflectivity has been attributed to phototransduction \cite{jonnal2007vivo}. This reflectivity can be changed as a response to a stimulus \cite{jonnal2007vivo, grieve2008intrinsic}. Indeed, it would be interesting to image the cone mosaic with FF-OCT during a light-evoked stimulus to study the reflectivity variation of each individual cone, a potential metric of photoreceptor health. FF-OCT seems a good imaging system to differentiate the IS/OS reflectivity signal of interest and multimodal signal from cone outer segment tip and non-cone sources (for example retinal pigment epithelium) \cite{litts2017photoreceptor}, thanks to the high axial resolution of the system.

\end{document}